\documentclass[twoside,slac_one]{revtex4}
\usepackage{graphicx}
\usepackage{fancyhdr}
\usepackage{amsmath} 
\usepackage{bm}
\usepackage{amsxtra}
\usepackage{amssymb}
\usepackage{amsthm}
\usepackage{latexsym}
\usepackage{lscape}

\newcommand{\tprods}[2]{\langle#1#2\rangle}

\newcommand{\beq}{\begin{equation}}
\newcommand{\eeq}{\end{equation}}
\newcommand{\beqs}{\begin{eqnarray}}
\newcommand{\eeqs}{\end{eqnarray}}
\newcommand{\Tr}{\ensuremath{\mathop{\mathrm{Tr}}}}

\def\be{\begin{equation}}
\def\ee{\end{equation}}
\def\bea{\begin{eqnarray}}
\def\eea{\end{eqnarray}}
\def\bsp{\be\begin{split}}
\def\la{\langle}
\def\ra{\rangle}

\def\a{\alpha}
\def\b{\beta}
\def\g{\gamma}

\def\d{\delta}

\def\m{\mu}

\def\r{\rho}
\def\l{\lambda}

\begin{document}

\title{Hidden Local and Non-local Symmetries of $S$ - matrices of $\mathcal{N}=2,4,8$ SYM in $D=2+1$}

%

\author{Abhishek Agarwal\footnote{Talk presented by Abhishek Agarwal} }
\affiliation{American Physical Society, Physical Review Letters, Ridge, New York, USA}
\author{Donovan Young }
\affiliation{Niels Bohr Institute, Blegdamsvej 17,  Copenhagen, Denmark}

\begin{abstract}
This talk, based principally on \cite{AA-DU-S}, is devoted to
properties of tree-level $S$-matrices of $\mathcal{N}=2,4,8$ SYM in
$D=2+1$. We'll discuss an on-shell formalism for three-dimensional
theories inspired by the spinor-helicity framework in four spacetime
dimensions. Our framework will be shown be to particularly well suited
for the extraction of hidden symmetries and algebraic structures that
the scattering amplitudes of the three-dimensional theories posses. In
particular we shall discuss the manifest $SO(\mathcal{N})$ symmetry of
the $S$-matrix to all orders in perturbation theory; a symmetry that
the Lagrangians of these theories do not have. After a brief
discussion of the ramification of the $SO(\mathcal{N})$ invariance to
the $D2-M2$ brane dualities, we shall  introduce an on-shell
superfield framework for three-dimensional theories  and end with a
surprising hint of the existence of non-local symmetries for the
$S$-matrix of the $\mathcal{N} = 8$ theory.
\end{abstract}

\maketitle

\thispagestyle{fancy}


\section{Introduction}
As emphasized in various presentations in this session of the meeting,
and summarized particularly succinctly in Marcus Spradlin's talk,
there has been an enormous amount of progress in our understanding of
$S$-matrices of Yang-Mills theories with high degrees of
supersymmetry, particularly in four spacetime dimensions.  For recent
reviews of the subject, see \cite{reviews}. In particular we have
learned that the gauge invariant $S$-matrices reveal many additional
symmetries and algebraic structures (Yangian invariance and dual
superconformal invariance in the case of four-dimensional theories to
name a few) which are difficult, if not impossible, to fathom using
the standard Lagrangian formulations of the corresponding gauge
theories.  It is perhaps fair to say, that  the wealth of insights
uncovered by much of the recent work on  $S$-matrices  seems
ostensibly  to depend on the underlying conformal invariance that the
four-dimensional theories posses, at least at tree level. Thus a
natural question to ask is how the lessons learned in $D=4$ generalize
when one gives up on conformal symmetries. Yang-Mills theories in
three spacetime dimensions seem to provide a fertile test bed to probe
this question. In $D=3$, one has control over the amount of
supersymmetry that one might want to impose on the gauge theory,
however, the fact that the Yang-Mills coupling constant is
dimension-full in $D=3$ implies that conformal symmetry is lost from
the get-go. Apart from providing a controlled departure from conformal
invariance, SYM theories in $D=2+1$ are of intrinsic interest from the
point of view of the gauge gravity duality, in particular from the
point of view of applying $D2$ and $M2$ brane systems to the study of
condensed matter physics. With all these motivations in mind, we'll
focus on the studies of $S$ matrices of the dimensional reductions of
$\mathcal{N}=1$ SYM from $D =$ 4, 6, and 10 to $D = 3$. 

We'll start with a pedagogical introduction to a spinor-helicity-like
on-shell formalism and use it to study the tree-level four particle
amplitudes of the $S$-matrix elements of the theories mentioned above
in a unified way. The point of this discussion will be to show that
the four particle amplitudes mirror the algebraic structures of the
corresponding quantities in four-dimensional theories in a close
way. Furthermore, the $S$-matrix will be shown to have full
$SO(\mathcal{N})$ invariance, even though the Lagrangian of the
corresponding theories only have manifest $SO(\mathcal{N}-1)$
symmetry. Afterwards, we'll recast our computations in an efficient
form using an on-shell superspace prescription and comment on possible
implications of the enhanced symmetry of the $S$-matrix to the connection
between $D2$ and $M2$ brane worldvolume theories. At the end of the
talk, we'll construct a non-local operator, reminiscent of 
a level-1 Yangian generator in four dimensions, that commutes with the
analogs of the MHV super-amplitudes in $D=3$, even in the absence of
dual-conformal and dual-superconformal symmetries.

\section{$SO(\mathcal{N})$ Covariant On-shell Framework}
We begin with a unified description of the Lagrangians of the theories under consideration; $\mathcal{N} = 2,4,8$ SYM in $D=2+1$, which are dimensional reductions of $\mathcal{N}=1$ SYM theories from $4,6,10$ dimensions respectively to three spacetime dimensions,
\be
 S = \frac{1}{g^2} \Tr \int d^3 x 
\Bigl(-\frac{1}{2} F_{\bar M\bar N} F^{\bar M \bar N} + i \bar \l_A
 \g^{\m} D_{\m} \l_A 
 +\r^i_{AB} \bar  \l_A [\Phi_i ,\l_B] \Bigr),\nonumber
\ee
where $F_{\bar M \bar N}$ is the field strength in 4,6 or 10 dimensions reduced to $D=2+1$.  $\bar M, \bar N$ are the three-dimensional Lorentz indices for $\bar M, \bar N = 0,1,2$.  For the other values of the barred indices the $F^2$ term is just a short-hand notation for the part of the action containing the $\mathcal{N}-1$ scalars  $\Phi ^i, \{i=2\cdots \mathcal{N}\}$ and their interactions with the gauge field. We also have $\mathcal{N}$ real fermions, $\lambda _A,  \{A=1\cdots \mathcal{N}\}$. From the part of the action involving the scalars, it is easily seen that these theories have a manifestly realized $SO(\mathcal{N}-1)$ R-symmetry at the Lagrangian level. 

Keeping the charge of the talk in mind, we can already see the emergence of a putative $SO(\mathcal{N})$ structure at the Lagrangian level. The tensors ($\rho $) that dictate the Yukawa couplings, when combined  with the obvious $SO(\mathcal{N})$ invariant, namely, the delta function satisfy the following equations:
\be
\rho ^C_{AB} = \{\rho^1_{AB} = \delta_{AB}, \rho^i_{AB}\},\qquad \r^D_{AC} \r^E_{BC} + \r^E_{AC} \r^D_{BC} =
2\d^{DE}\d_{AB}.
\ee
These tensors have natural  $SO(\mathcal{N})$ covariance properties. For $\mathcal{N}=2$,
\be
\rho^C_{AB} = \{\delta_{AB}, \epsilon_{AB}\}
\ee
are the two $SO(2)$ invariants. For $\mathcal{N}=8$, $\rho ^A_{BC}$ are the well known $\bf{8}_{s,c,v}$ symbols relating the three eight dimensional representations of $SO(8)$. As we shall see, the $\rho $ tensors are also the natural  structure constants that appear in the on-shell supersymmetry algebra.  To see that, we carry out an oscillator decomposition of the dynamical fields as follows

\bsp \label{modeexp}
&\Phi ^i =
\int \frac{d^2p}{(2\pi)^2}\frac{1}{\sqrt{2p^0}}\left(a^{i\dagger}(p)e^{ip\cdot x} +
a^i(p)e^{-ip\cdot x}\right),\\ &A_\mu = \int
\frac{d^2p}{(2\pi)^2}\frac{1}{\sqrt{2p^0}}\epsilon_{\mu}(p,k)\left(a_1^\dagger(p)e^{ip\cdot x}
+ a_1(p)e^{-ip\cdot x} \right),\\ 
& \lambda_I= \int
\frac{d^2p}{(2\pi)^2}\frac{1}{\sqrt{2p^0}}\left(u(p) \lambda_I^\dagger(p)e^{ip\cdot x} +
u(p)\lambda_I(p)e^{-ip\cdot x} \right).
\end{split}
\ee
Taking a cue from the spinor-helicity basis employed in four dimensions, we construct the polarization vector from the solutions of the massless Dirac equation as
\be \epsilon_\mu (p,k) =
\frac{\langle p|\gamma_\mu|k\rangle}{\tprods{k}{p}},\hspace{.3cm}
p_\mu \epsilon^\mu (p,k) = k_\mu \epsilon^\mu (p,k) = 0. 
\ee
It is
implied that
 \be |p\rangle = u(p), \hspace{.2cm} \langle p| =
\bar{u}(p), \hspace{.2cm} \tprods{k}{p} = \bar{u}(k)u(p) = -
\tprods{p}{k}, \label{prod1}\ee
 where $u(p)$ is a solution of the massless Dirac equation in three
 dimensions\footnote{$k$ is a reference momentum. The arbitrariness involved in the choice of $k$ is nothing but arbitrariness of gauge choices in disguise.}.
 
 Now, if we were studying a free abelian theory, then we would have been able to dualize the gauge field into a scalar $(\partial_\mu \Phi_1 \sim \epsilon_{\mu \nu \rho}F^{\nu \rho})$, which would have combined with the other $\mathcal{N}-1$ scalars to form an $SO(\mathcal{N})$ invariant combination. While we do not know how to dualize the non-abelian theory in an off-shell formalism, we can combine the on-shell scalar for the gauge field $a_i$ with the other $\mathcal{N}-1$ scalar oscillators $a_i$  to form an $SO(\mathcal{N})$ vector even in the non-abelian theory: $a_N = \{a_1, a_i\}$. As explained in \cite{AA-DU-S} we can take advantage of various properties of the polarization vector constructed above to deduce the supersymmetry transformation laws for the on-shell states to be the following:
\be
Q^\alpha _A |a_B(p)\rangle = \frac{1}{2}u^\alpha (p) \rho ^B_{AC}|\lambda _C(p)\rangle, \hspace{.2cm} Q^\alpha _A |\lambda_B(p)\rangle = -\frac{1}{2}u^\alpha(p)\rho^C_{AB}|a_C(p)\rangle.\label{susy}
\ee 
It is immediately clear from the appearance of the $SO(\mathcal{N})$ covariant quantities $\rho ^A_{BC}$ as the structure constants that the on-shell algebra is $SO(\mathcal{N})$, as opposed to $SO(\mathcal{N}-1)$, covariant. Thus for the $S$ matrix to commute with the supercharges, it must necessarily be $SO(\mathcal{N})$ invariant. This is a statement we will back up with some evidence later in the talk. For now let us focus on the enhanced symmetry of the on-shell framework.
\subsection{Symmetry Enhancement and Helicity}
The enhancement of symmetry from $SO(\mathcal{N}-1)$ to $SO(\mathcal{N})$ can be traced to $\mathcal{N}-1$ factors of $U(1)$ which relate the on-shell gluon $a_1$ to the oscillators corresponding to the scalar fields.  Focusing on the $\mathcal{N}=2$ case for simplicity; and defining
the complex combinations $\mathcal{W}_\pm =
\frac{1}{\sqrt{2}}(\mathcal{W}_1 \pm i \mathcal{W}_2)$, we can express
the algebra on single particle states in a $U(1)$ symmetric form as
\bsp\label{Q3} 
&Q^\alpha _+ |a_+\rangle = \frac{1}{\sqrt{2}} u^\alpha
|\lambda_+\rangle, \hspace{.4cm}Q^\alpha _+ |\lambda_-\rangle = -
\frac{1}{\sqrt{2}} u^\alpha |a_-\rangle,\\ &Q^\alpha _- |a_-\rangle =
\frac{1}{\sqrt{2}} u^\alpha |\lambda_-\rangle, \hspace{.4cm}Q^\alpha
_- |\lambda_+\rangle = - \frac{1}{\sqrt{2}} u^\alpha |a_+\rangle,\\
&Q^\alpha _- |a_+\rangle = Q^\alpha _+ |a_-\rangle = Q^\alpha _+
|\lambda_+\rangle = Q^\alpha _- |\lambda_-\rangle = 0.
\end{split}
\ee
The emergent $U(1)$ is nothing but what used to be the little group in $D=4$. Thus, the higher dimensional helicity degree of freedom augments itself as a continuous $U(1)$ symmetry at the level of the on-shell amplitudes upon dimensional reduction. As a matter of fact, it is shown in \cite{AA-DU-S} that it is precisely the dimensional reduction of the two gluon helicity states that generates the complex combination $a_\pm$ in $D=3$. In other words, the $U(1)$ charge carried by the one particle states plays exactly the same role that helicity does in four dimensions. For instance, using the supersymmetry algebra it may be shown that the color-ordered  amplitudes with `partons' with the same helicity, or with at most one $U(1)$ charge different, all vanish
\be
\langle ++\cdots ++\rangle = \langle --\cdots--\rangle = \langle \pm\pm\cdots \mp\pm\pm\rangle = 0.
\ee 
In a three-dimensional context, all of the above states can be shown to be $1/2$ BPS states.  The first non-vanishing amplitude is the one with two $U(1)$ charges different; the $D=3$ analog of the MHV amplitude. Quite remarkably it is given also by the famous Parke-Taylor formula
\be
\langle a_+ a_+\cdots a_-\cdots a_-\cdots a_+\rangle = \frac{2\la i j\ra^4}{\la12\ra\la23\ra\la34\ra \cdots \la n1\ra},
\ee
where $i,j$ are the momenta corresponding to the `-' excitations. Of course, the spinor products here are different than the four-dimensional ones, and are given by (\ref{prod1}). However, the algebraic structure of the amplitudes is exactly the same as in the four-dimensional case.
\subsection{Four Particle Amplitudes and $SO(\mathcal{N})$ Invariance }
Let us now focus on four particle amplitudes for the $\mathcal{N} = 2,4,8$ theories. There is only one independent four particle amplitude, which we can take to be the MHV amplitude given above. All others can be related to this amplitude by the action of the supersymmetry generators.  For instance:
\bsp\label{bbff} &\langle \lambda_+ \lambda_- a_+a_-\rangle =
+\frac{\tprods{3}{2}}{\tprods{3}{1}}\langle a_+a_-a_+a_-\rangle,\quad
\langle \lambda_+ \lambda_- a_-a_+\rangle =
+\frac{\tprods{4}{2}}{\tprods{4}{1}}\langle a_+a_-a_-a_+\rangle,\\
&\langle \lambda_+\lambda_-\lambda_-\lambda_+\rangle =
+\frac{\tprods{1}{4}}{\tprods{4}{2}}\langle
a_+a_-\lambda_-\lambda_+\rangle =
+\frac{\tprods{1}{3}}{\tprods{2}{4}}\langle a_+a_-a_-a_+\rangle.
\end{split}
\ee
All the relations between the different color-ordered four particle amplitudes of the $\mathcal{N}=2$ theory (which have also been explicitly checked by tree-level computations) are given in detail in \cite{AA-DU-S}. If we write the amplitudes in a $SO(\mathcal{N})$ symmetric form, then we can make the $SO(\mathcal{N})$ symmetry of these amplitudes manifest. To display a few classes of amplitudes:
\begin{eqnarray}
 \la a_{A_1} a_{A_2} a_{A_3} a_{A_4} \ra &=& -2\d_{A_1A_2}\d_{A_3A_4}\frac{\la 13\ra \la24\ra}{\la12\ra\la34\ra}
+ 2\d_{A_1A_3}\d_{A_2A_4}\nonumber\\
&&+ 2\d_{A_1A_4}\d_{A_2A_3}\frac{\la 13\ra \la24\ra}{\la23\ra\la41\ra}\nonumber
\end{eqnarray}
\begin{eqnarray}
\la \l_{A_1} \l_{A_2}\l_{A_3} \l_{A_4} \ra
&=& 2\d_{A_1A_2}\d_{A_3A_4}\frac{\la 13\ra^2 \la23\ra}{\la12\ra^2\la41\ra}
 -2\d_{A_1A_3}\d_{A_2A_4}\frac{\la34\ra}{\la12\ra}\nonumber\\
 &&-2\d_{A_1A_4}\d_{A_2A_3}\frac{\la 13\ra^2 \la12\ra}{\la14\ra^2\la34\ra}\nonumber
\end{eqnarray}
Defining $\r^{A_1A_2} \equiv \frac{1}{2}\Bigl((\r^{A_1})^T
\r^{A_2}-(\r^{A_2})^T \r^{A_1}\Bigr)$,
\be
\la a_{A_1} a_{A_2} \l_{A_3} \l_{A_4} \ra
= -\d_{A_1A_2}\d_{A_3A_4}\frac{\la 13\ra^2}{\la12\ra^2}\left(\frac{\la
  13\ra}{\la14\ra}+ \frac{\la 23\ra}{\la24\ra}\right)
+ \Bigl(\r^{A_1A_2}\Bigr)_{A_3A_4}\frac{\la31\ra}{\la14\ra}.\nonumber
\ee
The key point here is that all the amplitudes only involve the
$SO(\mathcal{N})$ invariant delta function which substantiates the
claim made earlier in the talk based on the on-shell superalgebra
i.e. the $S$-matrices of the theories under consideration have a
manifest $SO(\mathcal{N})$ symmetry, even though their Lagrangians
don't\footnote{We refer to \cite{AA-DU-S} for the explicit $SO(\mathcal{N})$ invariant forms of the remaining amplitudes.}.
\section{D2 vs M2}
One of the reasons to focus on the $SO(8)$ symmetry of the maximally supersymmetric $D=3$ SYM theory is to understand the flow of the theory to its infrared fixed point where it is expected to be described by a superconformal Chern-Simons-Matter (SCS) theory. In the case of the $SU(2)$ gauge group the conformal theory is expected to be the BLG theory \cite{blg}, which is of course a special case of the ABJM model \cite{abjm}. The conformal theory has a manifestly realized $SO(8)$ R-symmetry, while the SYM theory only has $SO(7)$ symmetry. One of the outstanding issues about the flow of the SYM theory had been to understand how the enhancement of the global symmetries takes place. However, as far as on-shell quantities are concerned, we have shown that the $SO(8)$ symmetry is manifestly realized to all orders in perturbation theory. Fortified by our results so far, we can begin to compare the structures of the four particle amplitudes of the two theories. 

All the four particle amplitudes of a large class of SCS theories with $\mathcal{N} \geq 4$ supersymmetry were computed in \cite{scs-scattering}. The amplitudes take on the form:
\be
S^{CS}(W_I, W_J, W_K, W_L) = S^{CS}(k,s,t) S_{IJKL}(s,t)
\ee
where the universal part $S_{IJKL}(s,t)$ is determined by the supersymmetry algebra alone.  $i,j,k,l$ stand for the flavor $SO(8)$ indices in the case of the BLG theory. All the dependence on the Chern-Simons level number (the coupling constant) is contained in the function $S^{CS}(k,s,t)$, which also depends on the Mandelstam variables. Further more, as far as the Yang-Mills amplitudes are concerned, we have already seen that they are of the form:
\be
S^{YM}(W_I, W_J, W_K, W_L) = S^{YM}(g^2,s,t) \tilde S_{IJKL}(s,t)
\ee
where the leading order term in $ S^{YM}(g^2,s,t)$ is given by the Parke-Taylor formula given before. Furthermore, using the $SO(8)$ covariant form of the SUSY algebra, it is easy to show \cite{AA-DU-S} that since both $ S_{IJKL}(s,t)$ and $\tilde  S_{IJKL}(s,t)$ follow from the same underlying algebra:
\be
 S_{IJKL}(s,t) =  \tilde S_{IJKL}(s,t).
 \ee
 Thus at the level of the four particle amplitudes,  the connection between $D2$ and $M2$ brane theories can be reduces to the problem of understanding the interpolating function which starts off as the Parke-Taylor formula and reduces to $S^{CS}(k,s,t)$ at infinite $g^2$. Given the rapidly growing body of exciting developments regarding  amplitudes of superconformal Chern-Simons theories \cite{scs-related}, this is certainly a worthwhile direction to follow for future research.
 \section{All Tree-Level Amplitudes via Superfields and a Potential Yangian-Like Symmetry}
 Moving beyond four particle amplitudes at the tree level, it is very convenient to introduce an on-shell superfield to aid computations. The superfield relevant to the maximally supersymmetric gauge theory can be regarded as a dimensional reduction of the $D = \mathcal{N} = 4$ superfield, which is  usually expressed as \cite{vpn-amplitude}:
 \be
\Phi = G^+ + \eta_a\lambda^a - \frac{1}{2!}\eta_a\eta_bS^{ab} -\frac{1}{3!}\eta_a\eta_b\eta_c\lambda^{abc} + \eta_1\eta_2\eta_3\eta_4G^-
\ee
where $\eta_i$ are anti-commuting variables carrying $SU(4)$ indices. Apart from the six scalars $S^{ab}$ one also has the two gluon polarization states, $G^\pm$, which from the three-dimensional sense are to be regarded as complex combinations of the on-shell gluon and the extra scalar obtained by the toroidal compactification of the fourth component of the gauge field.

One also has two $SU(2)$ indices $\alpha, \dot\alpha$ which are contracted according to the rules
\be
\tilde \lambda _{\dot \alpha} \tilde \omega ^{\dot \alpha} = \tprods{\tilde \lambda}{\tilde\omega}, \hspace{.3cm} \lambda _ \alpha \omega^\alpha = [\lambda \omega].
\ee

Dimensional reduction at the level of the superfield simply corresponds to identifying the two $SU(2)$ indices, which also leads to an identification of the two spinor products given above. For instance, The $D=4$ supercharges are:
\be
q^{\a a} = \l^\a \frac{\partial}{\partial \eta_a}, \hspace{.3cm} \bar q^{\dot \a}_a = \tilde\l ^{\dot \a}\eta _a,
\ee
which after dimensional reduction become 
\be
q^{\a a} = \l^\a \frac{\partial}{\partial \eta_a}, \hspace{.3cm} \bar q^{ \a}_a = \l ^{\a}\eta _a.
\ee
With these rules of dimensional reduction in place, we can (as in four dimensions) express all the three-dimensional MHV amplitudes as 
\be
\mathcal{A}^{mhv}_n = \frac{\delta ^3(p)\delta ^8(\tilde q)}{\tprods{1}{2}\tprods{2}{3}\cdots \tprods{n}{1}}, \hspace{.3cm} \delta ^8(\tilde q) = \frac{1}{2^4}\prod_{a=1}^4\sum_{i,j =1}^n\tprods{i}{j}\eta_{ia}\eta_{ja}.\label{mhvs}
\ee
More explicitly, all the MHV amplitudes with flipped $U(1)$ charges for the $i$ and $j$ entries  are given by\footnote{If one wants to work only with a $D=3$ theory with non-maximal supersymmetry, then the superfield methods of \cite{elvang} can be readily used in a three-dimensional context after they are subject to the same dimensional reduction mentioned above.}
\be
\mathcal{A}^{mhv}_n(i,j)  = \frac{1}{\tprods{1}{2}\tprods{2}{3}\cdots
  \tprods{n}{1}}\prod_{a=1}^4\left(\tprods{i}{j} + \tprods{i}{k}
\bar\eta^a_j\eta_{ka} - \tprods{j}{k}\bar\eta^a_i\eta_{ka} -
\frac{1}{2}\tprods{k}{l}\bar\eta^a_i\bar\eta^a_j\eta_{ka}\eta_{la}
\right)
\ee
where sum over repeated momentum indices is implied. $\bar \eta $ is understood to be the Fourier conjugate variable to $\eta$ in the sense that for a superfield $\Phi(\eta)$, the adjoint superfield is obtained by 
\[
\bar \Phi(\bar \eta) = \int d\eta_1d\eta_2d\eta_3d\eta_4e^{\eta_a\bar\eta^a}\Phi(\eta).
\] 

In \cite{AA-DU-S}, a precise  relation between the on-shell degrees of freedom of a four-dimensional gauge theory and those of the $D=3$ SYM theory obtained by its dimensional reduction were given. We can augment those relations by the on-shell dimensional reduction recipe given above at the level of the superfields. Now, a closed on-shell formula for $all$ tree-level amplitudes of $\mathcal{N}=4$ SYM in $D=4$ was given in \cite{all-n=4}. The simple procedure described above now allows us to use those results to get the answers for any desired tree-level amplitude in $D=3$ as well, albeit in a somewhat indirect form.

Finally, let us point out that using the superfield formalism one can
explicitly show that the $D=3$ tree amplitudes do $not$ posses
conformal or  dual-conformal symmetries. Yet, the tree-level theories
are as solvable as their four-dimensional counterparts (all the
amplitudes as we argue above, are known) which suggests that there
might be a Yangian-like non-local symmetry underlying the maximally
supersymmetric theory in three dimensions as well. In the absence of
dual conformal invariance, we cannot utilize the methods of \cite{DHP}
to probe the existence of Yangian symmetries or lack thereof in the
theories of interest to us. However, taking a more pedestrian
approach, we can construct a non-local charge that annihilates the
most symmetric of the three-dimensional amplitudes, namely the MHV
amplitudes given in (\ref{mhvs}).
Let us consider the following generator:
\be
\hat P^{\alpha \beta} = P^{\{\alpha \beta\}}\wedge {\cal D}+ P^{\delta \{\alpha}\wedge L^{\beta \}}_\delta + \bar{q}_c^{\{\beta}\wedge q^{c\alpha\}},
\ee
where $A\wedge B = \sum_{i<j}(A_iB_j - B_iA_j)$, the subscripts $i$
and $j$ label legs, and the curly brackets
denote symmetrization of indices. ${\cal D}$ and $L$ are the
dilatation and
the $SU(2)$ generators respectively, realized on-shell by:
\be\label{DandL}
{\cal D} = \frac{1}{2} \frac{\partial}{\partial \lambda ^\g }\lambda
^\g, \hspace{.3cm} L^\a _\b = \l^\a\frac{\partial}{\partial \lambda
  ^\b} - \frac{1}{2}\delta ^\a_\b \lambda ^\g
\frac{\partial}{\partial \lambda ^\g }.
\ee
A direct computation shows that 
\be
\hat P^{\alpha \beta} \mathcal{A}_n^{mhv} = 0.
\ee
The generator annihilating the amplitude is very reminiscent of the
level-1 bosonic Yangian generator for the tree level  $S$-matrix of
$\mathcal{N}=4$ SYM in $D=4$ \cite{DHP}, and SCS theories in
$D=3$ \cite{scs-related}. This result clearly hints at the possibility of
non-local symmetries in the $S$-matrix of three-dimensional maximally
supersymmetric Yang-Mills theory. As mentioned before, the absence of
dual conformal symmetries prevents us from generalizing this result to
non-MHV amplitudes. Perhaps a twistorial approach \cite{mini-twistor}
will help us determine if this non-local symmetry is just an artifact
of the MHV states or a true hidden symmetry of the $S$-matrix.

\section{Conclusions and Future Directions} 
To summarize, we have shown that amplitudes of $D=3$ SYM theories have additional symmetries which the corresponding Lagrangians do not have in a manifest form. In particular, we have argued (and explicitly demonstrated at the level of the four particle amplitudes) that the $S$-matrices of $\mathcal{N} = 2,4,8$ SYM theories are manifestly $SO(\mathcal{N})$ invariant, even though the actions only have  $SO(\mathcal{N}-1)$ R-symmetry. Furthermore, as seen by the Parke-Taylor form of the amplitudes in $D=3$, the explicit structures of the amplitudes have been shown to  mirror many of the properties of the amplitudes of the superconformal $\mathcal{N} = 4$ SYM in $D=4$, including a potential underlying non-local symmetry. The additional symmetries and the spinor-helicity framework presented in the talk is useful for studying the connections between $D2$ and $M2$ brane theories on-shell as well.

We would like to think of these results as an invitation to further explore properties of $S$-matrices of $D=3$ SYM theories. We are presently studying the nature of the loop-corrections and potential twistorial reformulations for the tree-level results presented here. We hope to report more results to complement  what has been presented here soon  
\begin{acknowledgments}
We would like to thank the organizers for arranging  a very enjoyable meeting and for giving us the opportunity to present  this work.
\end{acknowledgments}

\bigskip 

\end{document}